 \documentclass[final,5p,times,twocolumn]{elsarticle}

\usepackage{amssymb}

\journal{Physica C}

\begin{document}

\begin{frontmatter}



\title{Stability of the Kink State in a Stack of Intrinsic Josephson Junctions}


\author[add1,add2]{Shizeng Lin}
\author[add1,add2,add3]{and Xiao Hu}

\address[add1]{WPI Center for Materials Nanoarchitectonics, National Institute for Materials Science,Tsukuba 305-0044, Japan}
\address[add2]{Graduate School of Pure and Applied Sciences, University of Tsukuba, Tsukuba 305-8571, Japan}
\address[add3]{Japan Science and Technology Agency, 4-1-8 Honcho, Kawaguchi, Saitama 332-0012, Japan}

\begin{abstract}
A new dynamic state characterized by $(2m_l+1)\pi$ static phase kink with integers $\{m_l\}$ is proposed recently in a
stack of inductively coupled Josephson junctions. In the present paper, the stability of the phase kink state is
investigated against many perturbations and it is shown that the kink state is stable. It is also discussed that the
suppression of the amplitude of superconducting order parameter caused by the kink is weak.

\end{abstract}


\end{frontmatter}


\section{Introduction}
The experimental breakthrough\cite{Ozyuzer07,kadowaki08} on the terahertz radiation from a mesa of
$\rm{Bi_2Sr_2CaCu_2O_{8+\delta}}$(BSCCO) single crystal has triggered tremendous effort to reveal the mechanics of
emission. Soon after the experiments, it is proposed that a new dynamic state in a stack of inductively coupled
Josephson junctions is probably responsible for the experimental observations\cite{szlin08b, koshelev08b}. The new
dynamic state is characterized by $(2m_l+1)\pi$ phase kinks that are stacked along the $c$ direction and are localized
at the nodes of electric field, where $\{m_l\}$ are integers. To date, this kink state is the only solution that can
successfully explain the key observations in the experiments\cite{Ozyuzer07,kadowaki08}: first, the radiation frequency
and voltage obey the ac Josephson relation; secondly, strong and coherent emission appears at the cavity resonances. In
the present paper, we investigate the stability of the kink state.

\section{Results and discussion}
The dynamics of the gauge invariant phase difference in BSCCO is properly described by the inductively coupled
sine-Gordon equations\cite{Sakai93,Bulaevskii94}
\begin{equation}\label{eq1}
\partial_x^2P_l=(1-\zeta\Delta^{(2)})[\sin P_l+\partial_t^2P_l+\beta\partial_tP_l-J_{\rm{ext}}],
\end{equation}
where $P_l$ is the gauge invariant phase difference in the $l$th junction, $J_{\rm{ext}}$ the external current, $\beta$
the normalized conductance and $\zeta\sim10^5$ the inductive coupling. Dimensionless units are adopted and their
definitions are available in Ref. \cite{szlin09a}. Due to the presence of significant impedance mismatch as the case in
the experiments\cite{Ozyuzer07,kadowaki08}, the boundary condition may be approximated as $\partial_x P_l=0$ in the
absence of an external magnetic field.

The kink state as one solution to Eq. (\ref{eq1}) can be written as\cite{szlin08b,szlin09a,hu08}
\begin{equation}\label{eq2}
P_l(x,t)=\omega t + P_l^s(x) + {\mathop{\rm Re}\nolimits} [ - iA\cos ({k_1}x)\exp (i\omega t)],
\end{equation}
where the first is the rotating phase with voltage $\omega$, the second term the static phase kink and the last term
the plasma oscillation with $k_1\equiv\pi/L\gg 1$ with $L$ being the length of junction. Here we have considered the
voltage near the first cavity mode. Because of the huge inductive coupling, $P_l^s$ runs sharply from $0$ to
$(2m_l+1)\pi$ in the narrowed region of width $1/\sqrt{\zeta|A|}$. Approximating $P_l^s$ as a step function, we obtain
$A=4/\pi/(\omega^2-k_1^2-i\beta\omega)$. In the following, we consider the region where $A<1$.

To check the stability of the kink state Eq. (\ref{eq2}), we add a small perturbation to the solution,
$P_l'=P_l+\theta_l$ with $\theta_l\ll 1$. The kink state is stable if the perturbation dies out and is unstable if it
increases when $t\rightarrow+\infty$. Substituting $P_l'$ into Eq. (\ref{eq1}), we obtain the equation governing the
evolution of the perturbation
\begin{equation}\label{eq3}
\begin{array}{l}
 \partial _x^2{\theta _l} = (1 - \zeta {\Delta ^{(2)}})\{ [\frac{{\exp (i(\omega t + P_l^s)) + \exp ( - i(\omega t + P_l^s))}}{2} \\
  - \frac{A}{2}\cos ({k_1}x)\exp ( - iP_l^s)]{\theta _l} + \beta {\partial _t}{\theta _l} + \partial _t^2{\theta _l}\}.  \\
 \end{array}
\end{equation}
Here we investigate the region that $\omega \gg 1$, $A\ll 1$ and the variation of $\theta_l$ is much faster than that
of $P_l^s$ along the $c$ direction in the following calculation. The solution of $\theta_l$ can be expressed
as\cite{Bulaevskii07}
\begin{equation}\label{eq4}
\begin{array}{l}
 {\theta _l}(x,t) =  \\
 \sum\limits_q {\left[ {\overline {{\theta _q}}  + {\theta _{q + }}\exp (i\omega t) + {\theta _{q - }}\exp ( - i\omega t)} \right]\exp (iql)\exp ( - i\Omega t)}.  \\
 \end{array}
\end{equation}
Here the complex eigenfrequency $\Omega$ is assumed to be small $\Omega \ll 1$. The higher frequency harmonics $\Omega
\pm m \omega$ with $m>1$ in Eq. (\ref{eq4}) are small in the region of $\omega \gg 1$ and are neglected. Periodic
boundary condition is imposed along the $c$ direction, and hence $q=2m\pi/N$ with $m$ being an integer and $N$ the
total number of junctions. The kink state is stable if and only if $\rm{Im}{\Omega}<0$ for all $q$.

With the condition that $P_l^s$ has slow variation along the $c$ axis, we have
\begin{equation}\label{eq5}
(1 - \zeta {\Delta ^{(2)}})[\exp(iP_l^s)\theta_l]\thickapprox \exp(iP_l^s)\sum\limits_q a_q \theta_q \exp(iql),
\end{equation}
where $a_q\equiv 1+2\zeta (1-\cos q)$. Furthermore, because $P_l^s$ is almost a step function,
$\exp(iP_l^s)\thickapprox \cos P^s\equiv1-2\Theta(x-L/2)$ where$\Theta(x)$ is the Heaviside step function. Substituting
Eq. (\ref{eq4}) into Eq. (\ref{eq3}) and using Eq. (\ref{eq5}), we have for the frequency components $\omega\pm\Omega$
and $\Omega$
\begin{equation}\label{eq6}
{\partial _x^2\overline {{\theta _q}}  = {a_q}\left[ {\frac{{\cos {P^s}}}{2}({\theta _{q + }} + {\theta _{q - }}) -
\frac{{\cos {P^s}}}{2}A\cos ({k_1}x)\overline {{\theta _q}}  - \widetilde{{\Omega ^2}}\overline {{\theta _q}} }
\right]}
\end{equation}
\begin{equation}\label{eq7}
{\partial _x^2{\theta _{q \pm}} = {a_q}\left[ {\frac{{\cos {P^s}}}{2}\overline {{\theta _q}} - \widetilde{\omega _ \pm
^2}{\theta _{q \pm }}} \right]},
\end{equation}
where a small term proportional to $A$ in the coefficient of $\theta_{q \pm }$ in Eq. (\ref{eq7}) is neglected and
\begin{equation}\label{eq8}
\widetilde{{\Omega ^2}} \equiv {\Omega ^2} + \beta i\Omega ;\;\widetilde{\omega _ \pm ^2} \equiv {( \pm \omega  -
\Omega )^2} - \beta i( \pm \omega  - \Omega ).
\end{equation}
By inspecting Eq. (\ref{eq6}), the length scale over which $\overline {{\theta _q}}$ varies is of order of
$1/(\Omega\sqrt{a_q})\gg 1$. To solve Eq. (\ref{eq7}), we resort to multimodes expansion
\begin{equation}\label{eq9}
\cos {P^s} = \sum\limits_{n = 0} {{a_n}\cos ({k_n}x)} ,\;\;{\theta _{q \pm }} = \sum\limits_{n = 0} {{b_{n \pm }}\cos
({k_n}x)} ,
\end{equation}
with ${{a_n} = \frac{4}{{n\pi }}\sin (\frac{{n\pi }}{2})}$. Substituting Eq. (\ref{eq9}) into Eq. (\ref{eq7}) and
neglecting the coordinate dependence of $\overline {{\theta _q}}$, we obtain
\begin{equation}\label{eq10}
{b_{n \pm }} = \frac{{{a_n}\overline {{\theta _q}} {a_q}/2}}{{{a_q}\widetilde{\omega _ \pm ^2} - k_n^2}}.
\end{equation}
Inserting Eq.(\ref{eq9}) into Eq. (\ref{eq6}), we obtain the equation for $\overline {{\theta _q}}$
\begin{equation}\label{eq11}
\begin{array}{l}
 \partial _x^2\overline {{\theta _q}}  + {a_q}\widetilde{{\Omega ^2}}\overline {{\theta _q}}  \\
  = {a_q}\left[ {\frac{{\cos {P^s}}}{2}\sum\limits_n {({b_{n + }} + {b_{n - }})\cos ({k_n}x)}  - \frac{{\cos {P^s}}}{2}A\cos ({k_1}x)\overline {{\theta _q}} } \right]. \\
 \end{array}
\end{equation}
Equation (\ref{eq11}) and the boundary condition $\partial_x \overline {{\theta _q}}=0$ determine the spectrum of small
perturbations to the kink state. Integrating Eq.(\ref{eq11}) from $0$ to $L$ and approximating $\overline {{\theta
_q}}$ in the integrand as a constant when it is multiplied by other spatial dependent functions, we obtain the
dispersion relation
\begin{equation}\label{eq12}
{\Omega ^2} + \beta i\Omega  = \frac{1}{4}\sum\limits_{n = 0} {a_n^2\frac{{{\omega ^2} - {{(k'_n)}^2}}}{{{{[{\omega ^2}
- {{(k'_n)}^2}]}^2} + {\beta ^2}{\omega ^2}}}}  - \frac{{A{a_1}}}{4},
\end{equation}
with $k'_n\equiv k_n/\sqrt{a_q} \ll 1$. It is straightforward to check that $\rm{Im}{\Omega}<0$ when $\omega<k_1$, but
solution with $\rm{Im}{\Omega}>0$ appears when $\omega>k_1$. Thus we conclude that \emph{the kink state is stable in
the region of $\omega<k_1$}.

To study the full spectrum of stability, we resort to numerical simulation. It is found that the state with kink is
stable against different types of distortions, such as small perturbations both in the lateral direction and stack
direction, thermal fluctuations, weak external magnetic field\cite{szlin09a} and modulation of critical
current\cite{koshelev08b}. The radiation which is neglected in the analytical treatment is found to reduce the
stability of kink state to a certain degree\cite{szlin08b}.

Finally, we discuss the impact of the sharp phase change at the kink on the amplitude of superconducting order
parameter\cite{Tachiki09}. It is recognized that Eq. (\ref{eq1}) assumes the amplitude of superconductivity to be a
constant over space and time. To study fully the suppression of the superconductivity by phase kink, one needs to start
from equations self-consistently describing the evolution of phase as well as amplitude of superconductivity. However,
we assume the amplitude of superconductivity is fixed, which corresponds to infinite condensation energy limit, and
then consider the effect of phase kink.

The non-uniform alignment of kink along the $c$ axis induces in-plane magnetic field and supercurrent. The magnetic
field is $\sqrt{|A|/\zeta} \ll 1$ and the in-plane supercurrent is of order of $100\rm{J_c}$ with $J_c$ the critical
current along the $c$ axis. Compared with the critical magnetic field and critical current density, Both of them are
too small to suppress the superconductivity sensibly. From the energetic point of view, the energy costs caused by
phase kink is of the order of Josephson coupling, which is much smaller than the condensation energy in BSCCO. Thus, it
cannot suppress appreciably the amplitude of superconducting order parameter which is proportional to the condensation
energy. All these facts indicate the assumption of constant superconductivity amplitude is self-consistent and the
effect of phase kink on the superconductivity can be safely neglected. This estimate is consistent with the established
treatment on soliton phenomena in Josephson junctions\cite{fulton73}; although the suppression of superconductivity
amplitude has never been included in junction physics, calculated \emph{IV} curves for soliton states agree well with
experimental results. If one takes into account the suppression of superconductivity amplitude due to the phase kink,
the total energy is reduced from the one estimated by fixing the amplitude, which corresponds to the infinite limit of
superconductivity condensation.

In short, the kink state is stable and has negligibly small effect on the superconductivity.

\section{Acknowledgements}
This work was supported by WPI Initiative on Materials Nanoarchitronics, MEXT, Japan, CREST-JST, Japan and partially by
ITSNEM of CAS.



\begin{thebibliography}{10}
\expandafter\ifx\csname url\endcsname\relax
  \def\url#1{\texttt{#1}}\fi
\expandafter\ifx\csname urlprefix\endcsname\relax\def\urlprefix{URL }\fi \expandafter\ifx\csname href\endcsname\relax
  \def\href#1#2{#2} \def\path#1{#1}\fi

\bibitem{Ozyuzer07}
L.~Ozyuzer, A.~E. Koshelev, C.~Kurter, N.~Gopalsami, Q.~Li, M.~Tachiki,
  K.~Kadowaki, T.~Yamamoto, H.~Minami, H.~Yamaguchi, T.~Tachiki, K.~E. Gray,
  W.~K. Kwok, U.~Welp,
  Science 318 (2007) 1291.

\bibitem{kadowaki08}
K.~Kadowaki, H.~Yamaguchi, K.~Kawamata, T.~Yamamoto, H.~Minami, I.~Kakeya,
  U.~Welp, L.~Ozyuzer, A.~Koshelev, C.~Kurter, K.~Gray, W.-K. Kwok, Physca C
  468 (2008) 634.

\bibitem{szlin08b}
S.~Z. Lin, X.~Hu, Phys. Rev. Lett. 100 (2008) 247006.

\bibitem{koshelev08b}
A.~E. Koshelev, Phys. Rev. B 78 (2008) 174509.

\bibitem{Sakai93}
S.~Sakai, P.~Bodin, N.~F. Pedersen, J. Appl. Phys. 73 (1993)
  2411.

\bibitem{Bulaevskii94}
L.~N. Bulaevskii, M.~Zamora, D.~Baeriswyl, H.~Beck, J.~R. Clem, Phys. Rev. B 50 (1994) 12831.

\bibitem{szlin09a}
S.~Z. Lin, X.~Hu, Phys. Rev. B 79 (2009) 104507.

\bibitem{hu08}
X.~Hu, S.~Z. Lin, Phys. Rev. B 78 (2008) 134510.



\bibitem{Bulaevskii07}
L.~N. Bulaevskii, A.~E. Koshelev, Phys. Rev. Lett. 99 (2007) 057002.

\bibitem{Tachiki09}
M.~Tachiki, S.~Fukuya, T.~Koyama, Phys. Rev. Lett. 102 (2009) 127002.

\bibitem{fulton73}
{T. A. Fulton and R. C. Dynes}, Solid State Commun. 12 (1973) 57.

\end{thebibliography}

\end{document}